\documentclass[reprint,superscriptaddress,preprintnumbers,longbibliography,
amsmath,amssymb,aps,floatfix,pra,twocolumn, tightenlines 
]{revtex4-2}
\usepackage{subfigure}
\usepackage{mathrsfs}
\usepackage[T1]{fontenc}
\usepackage{graphicx}
\usepackage{dcolumn}
\usepackage{bm}
\usepackage{color}
\usepackage{times}
\usepackage{physics}
\usepackage[normalem]{ulem}
\usepackage{xurl}
\usepackage[colorlinks=true,
citecolor=blue,
linkcolor=magenta,
anchorcolor=black,
urlcolor=magenta]{hyperref}

\usepackage{soul}
\usepackage{xcolor} 
\usepackage{ulem}    
\usepackage{calc}    

\makeatletter
\def\NAT@spacechar{}
\makeatother

\usepackage{siunitx}
\usepackage{overpic}
\usepackage{braket}
\usepackage{changes}
\usepackage{setspace}
\usepackage{amssymb}

\usepackage{float}

\begin{document}

	\title{Observation and Modulation of the Quantum Mpemba Effect on a Superconducting Quantum Processor}
	
	\author{Yueshan Xu}
	\thanks{These authors contributed equally to this work.}
	\affiliation{Beijing Key Laboratory of Fault-Tolerant Quantum Computing, Beijing Academy of Quantum Information Sciences, Beijing 100193, China}
	
	\author{Cai-Ping Fang}
	\thanks{These authors contributed equally to this work.}
	\affiliation{Beijing National Laboratory for Condensed Matter Physics, Institute of Physics, Chinese Academy of Sciences, Beijing 100190, China}
	\affiliation{School of Physical Sciences, University of Chinese Academy of Sciences, Beijing 100049, China}
	
	\author{Bing-Jie Chen}	
	\thanks{These authors contributed equally to this work.}
	\affiliation{Beijing National Laboratory for Condensed Matter Physics, Institute of Physics, Chinese Academy of Sciences, Beijing 100190, China}
	\affiliation{School of Physical Sciences, University of Chinese Academy of Sciences, Beijing 100049, China}
	
	\author{Ming-Chuan Wang}
	\thanks{These authors contributed equally to this work.}
	\affiliation{Beijing National Laboratory for Condensed Matter Physics, Institute of Physics, Chinese Academy of Sciences, Beijing 100190, China}
	\affiliation{School of Physical Sciences, University of Chinese Academy of Sciences, Beijing 100049, China}
	
	\author{Zi-Yong Ge}
	\affiliation{Quantum Information Physics Theory Research Team, Center for Quantum Computing, RIKEN, Wako-shi, Saitama 351-0198, Japan}
	
	\author{Yun-Hao Shi}
	\affiliation{Beijing National Laboratory for Condensed Matter Physics, Institute of Physics, Chinese Academy of Sciences, Beijing 100190, China}

	\author{Yu Liu}
	\affiliation{Beijing National Laboratory for Condensed Matter Physics, Institute of Physics, Chinese Academy of Sciences, Beijing 100190, China}
	
	\author{Cheng-Lin Deng}
	\affiliation{Beijing Key Laboratory of Fault-Tolerant Quantum Computing, Beijing Academy of Quantum Information Sciences, Beijing 100193, China}
	
	\author{Kui Zhao}
	\affiliation{Beijing Key Laboratory of Fault-Tolerant Quantum Computing, Beijing Academy of Quantum Information Sciences, Beijing 100193, China}

	\author{Zheng-He Liu}
	\affiliation{Beijing National Laboratory for Condensed Matter Physics, Institute of Physics, Chinese Academy of Sciences, Beijing 100190, China}
	\affiliation{School of Physical Sciences, University of Chinese Academy of Sciences, Beijing 100049, China}

	\author{Tian-Ming Li}
	\affiliation{Beijing National Laboratory for Condensed Matter Physics, Institute of Physics, Chinese Academy of Sciences, Beijing 100190, China}
	\affiliation{School of Physical Sciences, University of Chinese Academy of Sciences, Beijing 100049, China}
	
	\author{Hao Li}
	\affiliation{Beijing Key Laboratory of Fault-Tolerant Quantum Computing, Beijing Academy of Quantum Information Sciences, Beijing 100193, China}
	
	\author{Ziting Wang}
	\affiliation{Beijing Key Laboratory of Fault-Tolerant Quantum Computing, Beijing Academy of Quantum Information Sciences, Beijing 100193, China}
	
	\author{Gui-Han Liang}
	\affiliation{Beijing National Laboratory for Condensed Matter Physics, Institute of Physics, Chinese Academy of Sciences, Beijing 100190, China}

	\author{Da'er Feng}
	\affiliation{Beijing National Laboratory for Condensed Matter Physics, Institute of Physics, Chinese Academy of Sciences, Beijing 100190, China}
	\affiliation{School of Physical Sciences, University of Chinese Academy of Sciences, Beijing 100049, China}
	
	\author{Xue-Yi Guo}
	\affiliation{Beijing Key Laboratory of Fault-Tolerant Quantum Computing, Beijing Academy of Quantum Information Sciences, Beijing 100193, China}
	
	\author{Xu-Yang Gu}
	\affiliation{Beijing National Laboratory for Condensed Matter Physics, Institute of Physics, Chinese Academy of Sciences, Beijing 100190, China}
	\affiliation{School of Physical Sciences, University of Chinese Academy of Sciences, Beijing 100049, China}
	
	\author{Yang He}
	\affiliation{Beijing National Laboratory for Condensed Matter Physics, Institute of Physics, Chinese Academy of Sciences, Beijing 100190, China}
	\affiliation{School of Physical Sciences, University of Chinese Academy of Sciences, Beijing 100049, China}
	
	\author{Hao-Tian Liu}
	\affiliation{Beijing National Laboratory for Condensed Matter Physics, Institute of Physics, Chinese Academy of Sciences, Beijing 100190, China}
	\affiliation{School of Physical Sciences, University of Chinese Academy of Sciences, Beijing 100049, China}

	\author{Zheng-Yang Mei}
	\affiliation{Beijing National Laboratory for Condensed Matter Physics, Institute of Physics, Chinese Academy of Sciences, Beijing 100190, China}
	\affiliation{School of Physical Sciences, University of Chinese Academy of Sciences, Beijing 100049, China}

	\author{Yongxi Xiao}
	\affiliation{Beijing National Laboratory for Condensed Matter Physics, Institute of Physics, Chinese Academy of Sciences, Beijing 100190, China}
	\affiliation{School of Physical Sciences, University of Chinese Academy of Sciences, Beijing 100049, China}

	\author{Yu Yan}
	\affiliation{Beijing National Laboratory for Condensed Matter Physics, Institute of Physics, Chinese Academy of Sciences, Beijing 100190, China}
	\affiliation{School of Physics, Northwest University, Xi’an 710127, China}
	
	\author{Yi-Han Yu}
	\affiliation{Beijing National Laboratory for Condensed Matter Physics, Institute of Physics, Chinese Academy of Sciences, Beijing 100190, China}
	\affiliation{School of Physical Sciences, University of Chinese Academy of Sciences, Beijing 100049, China}
	
	\author{Wei-Ping Yuan}
	\affiliation{Beijing National Laboratory for Condensed Matter Physics, Institute of Physics, Chinese Academy of Sciences, Beijing 100190, China}
	\affiliation{School of Physical Sciences, University of Chinese Academy of Sciences, Beijing 100049, China}
	
	\author{Jia-Chi Zhang}
	\affiliation{Beijing National Laboratory for Condensed Matter Physics, Institute of Physics, Chinese Academy of Sciences, Beijing 100190, China}
	\affiliation{School of Physical Sciences, University of Chinese Academy of Sciences, Beijing 100049, China}

    \author{Zheng-An Wang}
    \affiliation{Beijing Key Laboratory of Fault-Tolerant Quantum Computing, Beijing Academy of Quantum Information Sciences, Beijing 100193, China}
	
	\author{Gangqin Liu}
	\affiliation{Beijing National Laboratory for Condensed Matter Physics, Institute of Physics, Chinese Academy of Sciences, Beijing 100190, China}
	
	\author{Xiaohui Song}
	\affiliation{Beijing National Laboratory for Condensed Matter Physics, Institute of Physics, Chinese Academy of Sciences, Beijing 100190, China}
	
	\author{Ye Tian}
	\affiliation{Beijing National Laboratory for Condensed Matter Physics, Institute of Physics, Chinese Academy of Sciences, Beijing 100190, China}

	\author{Yu-Ran Zhang}
	\affiliation{School of Physics and Optoelectronics, South China University of Technology, Guangzhou 510640, China}

	\author{Shi-Xin Zhang}
	\affiliation{Beijing National Laboratory for Condensed Matter Physics, Institute of Physics, Chinese Academy of Sciences, Beijing 100190, China}
	
	\author{Kaixuan Huang}
	\email{huangkx@baqis.ac.cn}
	\affiliation{Beijing Key Laboratory of Fault-Tolerant Quantum Computing, Beijing Academy of Quantum Information Sciences, Beijing 100193, China}
	
	\author{Zhongcheng Xiang}
	\email{zcxiang@iphy.ac.cn}
	\affiliation{Beijing National Laboratory for Condensed Matter Physics, Institute of Physics, Chinese Academy of Sciences, Beijing 100190, China}
	\affiliation{School of Physical Sciences, University of Chinese Academy of Sciences, Beijing 100049, China}
	\affiliation{Hefei National Laboratory, Hefei 230088, China}
	
	\author{Dongning Zheng}
	\affiliation{Beijing National Laboratory for Condensed Matter Physics, Institute of Physics, Chinese Academy of Sciences, Beijing 100190, China}
	\affiliation{School of Physical Sciences, University of Chinese Academy of Sciences, Beijing 100049, China}
	\affiliation{Hefei National Laboratory, Hefei 230088, China}
	\affiliation{Songshan Lake Materials Laboratory, Dongguan, Guangdong 523808, China}
	
	\author{Kai Xu}
	\email{kaixu@iphy.ac.cn}
	\affiliation{Beijing National Laboratory for Condensed Matter Physics, Institute of Physics, Chinese Academy of Sciences, Beijing 100190, China}
	\affiliation{School of Physical Sciences, University of Chinese Academy of Sciences, Beijing 100049, China}
    \affiliation{Beijing Key Laboratory of Fault-Tolerant Quantum Computing, Beijing Academy of Quantum Information Sciences, Beijing 100193, China}
	\affiliation{Hefei National Laboratory, Hefei 230088, China}
	\affiliation{Songshan Lake Materials Laboratory, Dongguan, Guangdong 523808, China}
	
	\author{Heng Fan}
	\email{hfan@iphy.ac.cn}
	\affiliation{Beijing National Laboratory for Condensed Matter Physics, Institute of Physics, Chinese Academy of Sciences, Beijing 100190, China}
	\affiliation{School of Physical Sciences, University of Chinese Academy of Sciences, Beijing 100049, China}
    \affiliation{Beijing Key Laboratory of Fault-Tolerant Quantum Computing, Beijing Academy of Quantum Information Sciences, Beijing 100193, China}
	\affiliation{Hefei National Laboratory, Hefei 230088, China}
	\affiliation{Songshan Lake Materials Laboratory, Dongguan, Guangdong 523808, China}

	
	\begin{abstract}
    In non-equilibrium quantum systems, the quantum Mpemba effect (QME) emerges as a counterintuitive phenomenon: systems exhibiting greater initial symmetry breaking restore symmetry faster. It has been attracting broad interest in studying QME dynamics and potential applications in quantum information science. While theoretical exploration of QME has surged, experimental studies, specifically on its flexible modulation, remain limited. 
    Here, we report the observation and modulation of QME using a superconducting processor featuring an all-to-all connected, tunable-coupling architecture that enables precise control from short- to long-range interactions.
    This platform allows independent manipulation of coupling regimes, on-site potentials, and initial states, enabling us to elucidate their roles in QME. To quantify symmetry restoration, we employ entanglement asymmetry (EA), derived from the reconstructed density matrix via quantum state tomography, as a sensitive probe. In strong short-range coupling regimes, EA crossovers during quenches from tilted Néel states confirm the presence of QME. In contrast, in intermediate coupling regimes, synchronized EA and entanglement entropy dynamics reveal the suppression of QME. Remarkably, QME reemerges with the introduction of on-site linear potentials or quenches from tilted ferromagnetic states, the latter proving robust against on-site disorder. Our study demonstrates flexible QME modulation on a superconducting platform with multiple controllable parameters, shedding light on quantum many-body non-equilibrium dynamics and opening avenues for quantum information applications.

	\end{abstract}
	
	\maketitle

\textit{Introduction}---Relaxation to equilibrium is a ubiquitous and fundamental process in nature. In the quantum realm, understanding and manipulating this process is key not only for exploring thermalization and ergodicity-breaking theoretically~\cite{RMP_Nonequilibrium,RMP_thermalization}, but also for practical tasks such as efficient quantum state preparation and qubit reset~\cite{2022_NRP_information_science,PRL_PreparationStates,PhysRevX_2024,Gonzales2025_state,Aamir2025_resets,PRL_Reset}. While systems generally relax monotonically to equilibrium, some remarkable exceptions exist. Among these, the Mpemba effect stands out as a counterintuitive phenomenon, famously observed in the faster freezing of initially hotter water~\cite{Mpemba_effect} and later found across a wide range of classical systems~\cite{ME_index_prx,Nature2020_colloidal,2024_PRL,Thermomajorization_ME}. 
Recently, the Mpemba effect has been extended into the quantum regime~\cite{ME_review_classical}, manifesting in two distinct forms: (i) in open quantum systems interacting with an external environment through Markovian~\cite{QNQS_prl,Batteries_QME_2025,IQME_2025,dissipative_dynamics_all_connect,eigenvalue_crossing,NOQS_prr,heat_engines_PRA,open_mark_dot_PRB,PhysRevE_2023_markov,2024_OL,light_ME} and non-Markovian processes~\cite{Non_Markovian_QME,non_mark,non_mark_PRA,beyond_mark_2024} or (ii) in isolated quantum systems governed by intrinsic quantum dynamics. In open quantum systems, the relaxation processes associated with quantum strong Mpemba effect~\cite{strong_ME_nc_exp,strong_ME_exponential_acceleration_prl,strong_ME_exponential_acceleration_pra,strong_ME_exponential_acceleration_prr,ME_neq_free_energy,PhysRevE_2025_marko}, inverse Mpemba effect~\cite{inverse_ME_ion}, and related variants~\cite{QME_reservoirs_quantum_dot,2024_DMP,PhysRevA_MQME,2024superconducting_QME} are often dominated by classical fluctuations. By contrast, the relaxation dynamics in isolated systems are driven by intrinsic quantum fluctuations, offering a unique perspective on non-equilibrium behavior.

Although isolated quantum systems undergo unitary evolution, a local subsystem can experience the remainder of the system as an effective bath. In thermalizing regimes, the subsystem relaxes toward a statistical ensemble that respects the symmetry of the Hamiltonian, whereas integrable limits instead approach generalized Gibbs ensembles. In this context, the quantum Mpemba effect (QME) manifests as a striking phenomenon~\cite{EA_probe,NRP2025,QME_symm_presp_2025}: subsystems with greater initial symmetry breaking---the quantum analogue of hotter water in the classical Mpemba effect---restore the symmetry faster. This counterintuitive dynamics has been predicted across diverse quantum systems, particularly within the quasiparticle framework~\cite{Integrable_PRL,quasiparticle,integrable_arxiv} for integrable systems that include one-dimensional~\cite{XY_chain,XX_chain,mixed_states_dephase_prb,z2_XY_chain_2024,Heisenberg_chain,Rule54} and two-dimensional models~\cite{2d_free_fermion,optical_pra}.
Beyond these, the QME has been theoretically explored in chaotic or many-body localized (MBL) systems~\cite{random_circuits_2024,random_circuits_2025,QME_MBL,Translation_symmetry_random_2025,dual_unitary_circuits_PRXQ}, further revealing that its emergence depends on the initial state and ergodicity of systems. Experimental demonstrations, such as in trapped-ion systems with tilted ferromagnetic initial states~\cite{QME_trapped_ion}, confirm QME robustness against decoherence.
These pioneering studies indicate the potential generality and resilience of QME. Nevertheless, a comprehensive experimental investigation that systematically explores how factors such as interaction range, on-site potential and initial-state choice govern the QME remains an open challenge.

 Here, we address this gap using a novel fully connected superconducting quantum processor. By integrating a frequency-adjustable central bus resonator with tunable couplers, we achieve precise and flexible Hamiltonian engineering, enabling the exploration of symmetry-breaking dynamics with both tunable short- and long-range interactions. 
 To study the QME, we employ entanglement asymmetry (EA) as a sensitive probe of symmetry breaking~\cite{EA_probe}. 
 By tracking its evolution under a \(U(1)\)-symmetric Hamiltonian, we identify the characteristic EA crossover associated with the QME.
 
 In this work, we achieve observation and modulation of the QME by tuning the coupling strength regimes, applying on-site potentials, and preparing tilted Néel and ferromagnetic initial states. Specifically: (i) In the strong short-range limit, by quenching from tilted Néel states, we observe a clear crossing of EA curves, confirming that states with greater initial symmetry breaking restore symmetry faster—a hallmark of QME; (ii) In the intermediate coupling regime, EA decays monotonically without crossing, indicating the suppression of the QME in a thermalizing regime; (iii) In particular, we achieve reemergence of the QME through two independent mechanisms: Introducing static on-site potentials slows global thermalization, with larger tilt angles experiencing less suppression of thermalization, recovering an EA crossover; Alternatively, quenching from tilted ferromagnetic states maps the system to the Lipkin-Meshkov-Glick (LMG) model, restoring the QME with robust EA crossovers that persist under strong on-site disorder. These results reveal universal control knobs for non-equilibrium dynamics—interaction range, potential engineering, and initial state selection—offering new insights into many-body quantum dynamics. 

     \begin{figure*}[htbp]
         \centering
     	\includegraphics[width=0.93\linewidth]{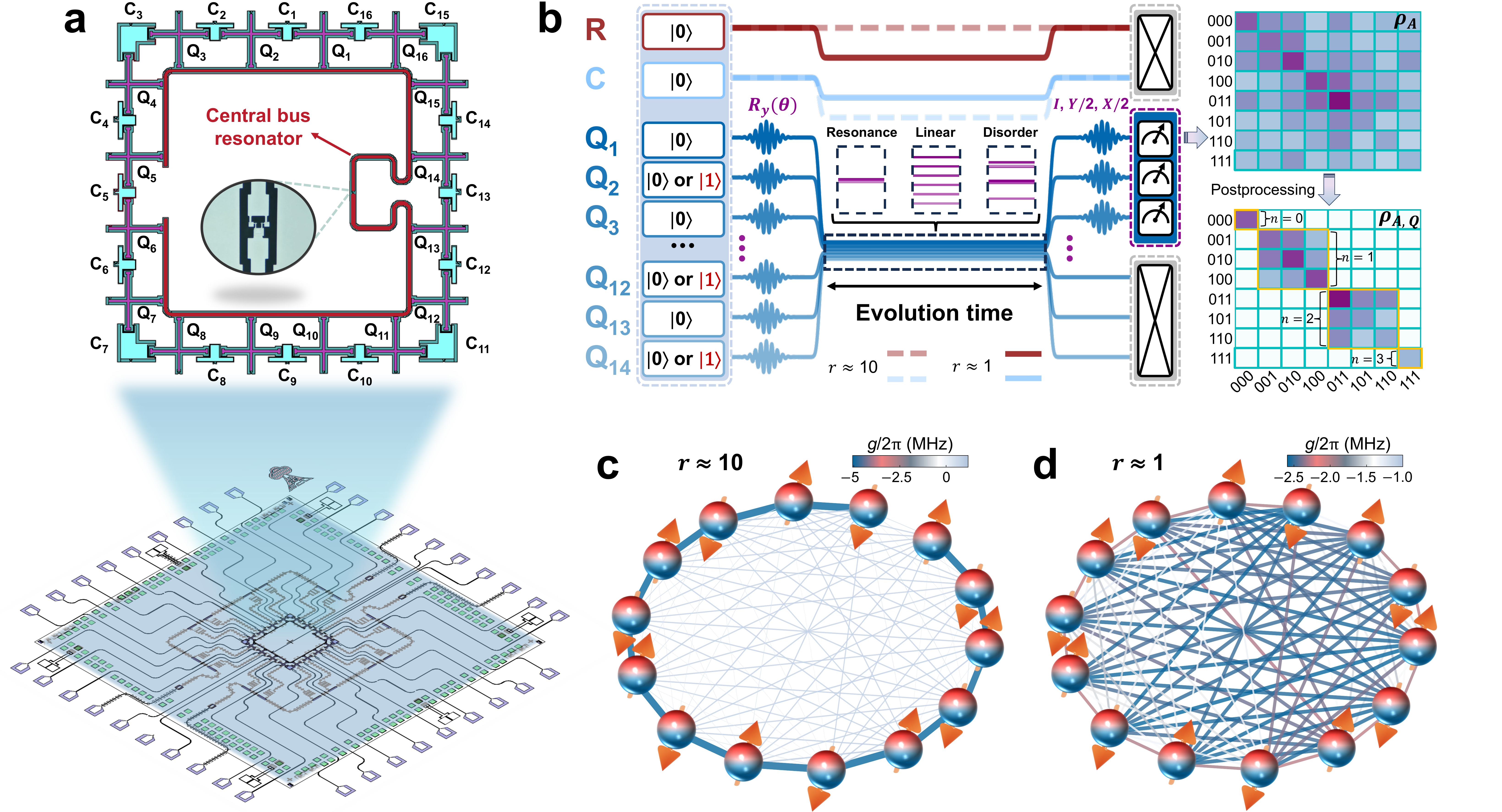}
     	\caption{Micrograph and coupling-strength modulation of the quantum processor. (a) Optical micrograph of the superconducting quantum processor, arranged in a ring topology with 16 qubits interconnected by tunable couplers ($C$) and a central bus resonator ($R$). The inset shows an optical micrograph of the Josephson junction at the center of the resonator. (b) Pulse sequences of the experimental protocol, which enable flexible modulation of coupling strengths via both $R$ and $C$. The protocol involves initializing the system in tilted product states, tuning qubits with various on-site potentials (resonance, disorder, or linear potentials), and constructing the density matrices of subsystem $ A = \{Q_1, Q_2, Q_3\} $ via QST. Following projection of the density matrix onto the eigenspaces of the charge operator $ Q_A $, classical postprocessing is applied to estimate the EA. The charge sectors are labeled by the eigenvalues $n$ of the excitation-number operator $\hat{n} = \sum_{i \in A} (1 - \sigma_i^z)/2$. (c, d) Schematics of the effective coupling strengths for $ r \approx 10 $ and $ r \approx 1 $, where $ r $ is the ratio of nearest-neighbor to long-range coupling strengths. Line thickness and color denote the magnitude.} 
	\label{fig1}
     \end{figure*}
     
	\textit{Experimental setup}---Superconducting quantum circuits~\cite{2019_science,2024_science,2024_nature}, with tunable interactions and high-fidelity control, provide an ideal platform for systematic QME studies. As illustrated in Fig.~\ref{fig1}(a), our 16-qubit superconducting quantum processor features a fully connected ring topology, with the capability of tunable qubit-qubit couplings in both short-range and long-range regimes. Therein, each qubit is capacitively coupled to its nearest-neighbor coupler, enabling precise dynamic control of short-range coupling strengths. In addition, all qubits are capacitively coupled to a central frequency-tunable bus resonator, establishing an effective all-to-all connectivity graph that facilitates programmable long-range interactions. In our experiments, we select 14 adjacent qubits arranged in a tightly connected chain array (see Supplemental Material for device parameters~\cite{supp_cite})\nocite{PRXQuantum_wyy,PRA_ltm,yan_coupler_2018,10_Qubit_2017,introduction_XYchain,NRP2025,Integrable_PRL,Xiang2023,QME_trapped_ion}. The qubits are labeled as \( Q_j \) for \(j = 1, 2, \ldots, 14 \), with the central bus resonator and couplers denoted as \( R \) and \( C \), respectively. 
	
	Following meticulous timing alignment and Z-pulse calibration~\cite{supp_cite}, we dynamically tune the frequencies of the qubits, couplers, and central bus resonator. The multiple tunable degrees of freedom of our device enables versatile control over the system’s Hamiltonian. With $\hbar$ settling to 1, the effective Hamiltonian of the system is expressed as~\cite{2021_PRL_guo}
    \begin{equation}\label{hamiltonian}
    \begin{aligned}
        H &= \sum_{i<j} g_{ij} (\sigma_i^+ \sigma_j^- + \sigma_i^- \sigma_j^+ ) + \sum_i h_i \sigma_i^+ \sigma_i^- \\
        &= \sum_{i,j=i+1} g_N (\sigma_i^+ \sigma_j^- + \sigma_i^- \sigma_j^+ ) \\
        &+ \sum_{i < j, j \neq i+1} g_L (\sigma_i^+ \sigma_j^- + \sigma_i^- \sigma_j^+)
        + \sum_i h_i \sigma_i^+ \sigma_i^-,
    \end{aligned} 
    \end{equation}
    where \(g_{ij}/2\pi\) is the coupling strength between \(Q_i\) and \(Q_j\), \(\sigma_i^+\) (\(\sigma_i^-\)) is the raising (lowering) operator, and \(h_{i}/2\pi\) denotes the on-site potential implemented through position-dependent frequency modulation, with a reference frequency \(\omega_{\mathrm{ref}}/2\pi \approx\) 4.24~GHz. In particular, we divide the coupling strength between qubits into two parts, yielding the short-range (nearest-neighbor) coupling strength \(g_N/2\pi\) and the long-range coupling strength \(g_L/2\pi\).  
    
The coupling ratio is defined as \( r = |g_N / g_L| \), which is tuned by adjusting the frequency detuning of the central bus resonator \( \Delta_{RQ} \) and couplers \( \Delta_{CQ} \) relative to the reference frequency. The variable \( r \) quantifies the different interaction regimes, spanning from strong short-range coupling regime (\( r \gg 1 \)), through an intermediate coupling regime (\( r \approx 1 \)), to weak short-range coupling regime (\( r \ll 1 \)). By adjusting the waveform sequence shown in Fig.~\ref{fig1}(b), a strong short-range coupling regime \( r \approx 10 \) (Fig.~\ref{fig1}(c)) is achieved when \( \Delta_{CQ} \approx 1.0 \, \text{GHz} \) and \( \Delta_{RQ} \approx 2.2 \, \text{GHz} \), while an intermediate coupling regime \( r \approx 1 \) (Fig.~\ref{fig1}(d)) is realized when \( \Delta_{CQ} \approx 2.0 \, \text{GHz} \) and \( \Delta_{RQ} \approx 0.4 \, \text{GHz} \). This tunability provides a novel degree of freedom for systematically investigating the dependence of the QME on different coupling regime.
    
We prepare tilted product states that break global $ U(1) $ symmetry by an angle $ \theta $, with qubit idle frequencies optimized for high-fidelity initial state preparation and minimal microwave crosstalk~\cite{supp_cite}.
The system is then subjected to a sudden quench under the engineered Hamiltonian $ H $, which incorporates tunable on-site potentials (resonance, disorder, or linear potentials) applied to the qubits (see Fig.~\ref{fig1}(b)). 
To explore the QME, we quantify the dynamics of symmetry breaking through EA, defined as the relative entropy
    \begin{equation}\label{EA}
\Delta S_A(t) = S(\rho_{A,Q}(t)) - S(\rho_A(t)),
    \end{equation}
where \(\rho_A \) is the reduced density matrix of subsystem \(A\) reconstructed via quantum state tomography (QST),  \(S(\rho_A)\) is the von Neumann (entanglement) entropy of the subsystem, and \(\rho_{A,Q} = \sum_q \Pi_q \rho_A \Pi_q\) is the projection of \(\rho_A\) onto eigenspaces of the conserved charge \(Q_A = \sum_{i \in A} \sigma_i^z\) (see Fig.~\ref{fig1}(b)). \(\Delta S_A(t)\) captures the contributions of quantum coherence between different charge sectors that violate the symmetry. In other words, EA quantifies the breaking of a given symmetry and vanishes only when \(\rho_A\) is already diagonal in the charge basis. This makes it a natural order parameter for non-equilibrium dynamics, offering a fresh perspective on thermalization compared to traditional metrics like entanglement entropy (EE).
Through the analysis of EA dynamics, we quantify the relaxation rates of states with different initial symmetry-breaking.

	\textit{QME in strong short-range coupling regime}---We initially explore the dynamics of EA in the strong short-range coupling regime ($ r \approx 10 $). The nearest-neighbor couplings are fixed at $ g_N/2\pi \approx -5 \, \text{MHz} $, while long-range interactions yield an average coupling $ \bar{g_L}/2\pi \approx 0.5 \, \text{MHz} $~\cite{supp_cite}, facilitating an approximate mapping to an integrable one-dimensional XX spin chain with open boundaries during early-time dynamics.

    The initial states are prepared as tilted Néel states, denoted $ \ket{\theta}_N $, with different tilt angles $ \theta=\pi/4 $ and $ \theta = \pi/2 $, constructed by applying a Y-rotation of angle $ \theta $ to each qubit. The state can be expressed as \( \ket{\theta}_N = \bigotimes_{j=1}^{14} R_y(\theta) \ket{s_j} \), where \( \ket{s_j} = \ket{0} \) ($\ket{1}$) for odd (even) \( j \), and the Y-rotation operator is defined as \( R_y(\theta) = e^{-i \theta \sigma_y / 2} \). The tilt angle \( \theta \) governs the degree of global \( U(1) \) symmetry breaking: in the Bloch sphere representation, \(\theta=0\) or \(\pi\) yields all spins aligned along \( \sigma_z\) (fully symmetric), whereas \(\theta=\pi/2\) aligns all spins along \(\sigma_x\) (maximal symmetry breaking), indicating \( \ket{\theta= \pi/2}_N \) exhibits greater initial symmetry breaking compared to \( \ket{\theta= \pi/4}_N \). In this sense, the state \( \ket{\theta= \pi/2}_N \) can be framed as the quantum analogue (not equivalence) of a higher initial temperature in the classical Mpemba effect.
    
     \begin{figure*}[htbp]
         \centering
     	\includegraphics[width=0.9\linewidth]{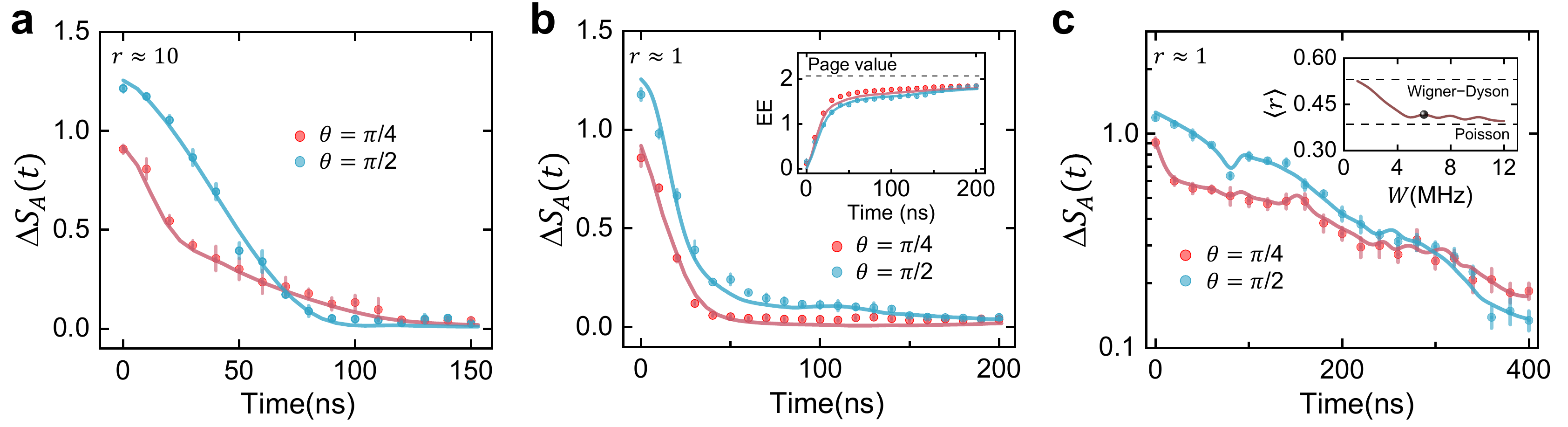}
     	\caption{EA dynamics in strong short-range coupling regime ($ r \approx 10 $) and intermediate coupling regime ($ r \approx 1 $) for two tilted Néel states. 
     	(a) EA dynamics under $ r \approx 10 $ with qubits on resonance ($h_i/2\pi = 0$). An obvious crossover is observed, confirming the presence of QME. 
     	(b) EA dynamics with qubits on resonance ($h_i/2\pi = 0$) for $ r \approx 1 $. The disappearance of the dynamical crossover, indicating the suppression of QME. The inset shows that the entanglement entropy (EE) approaches the Page value as time increases.
     	(c) EA dynamics in the presence of on-site potentials with \( W = 6 \, \text{MHz} \) for $ r \approx 1 $. Faster symmetry restoration for $\theta = \pi/2$, confirming the reemergence of QME. Symbols are experimental data and solid curves represent the theoretical results including decoherence. Error bars represent the standard deviation obtained from 10 experimental repetitions, each comprising 3000 shots. The inset shows that \(\langle r \rangle\) decreases as \( W \) increases.}

	\label{fig2}
     \end{figure*}     
    
    Following state preparation, we modulate the system to the target Hamiltonian with all qubits tuned to resonance, and track the EA dynamics for a terminal subsystem $A$. As depicted in Fig.~\ref{fig2}(a), despite the EA for $ \ket{\theta= \pi/2}_N $ initially exceeding that of $ \ket{\theta= \pi/4}_N $, the faster symmetry restoration of $ \ket{\theta= \pi/2}_N $ results in a pronounced crossover phenomenon—a hallmark of the QME. This effect originates from quasiparticle-mediated relaxation~\cite{Integrable_PRL,quasiparticle,integrable_arxiv}, wherein entangled quasiparticle pairs generated within subsystem $ A $ govern $ \Delta S_A $, with larger tilt angles $ \theta $ preferentially exciting faster-propagating modes, thereby accelerating symmetry restoration. Furthermore, our simulation results reveal the emergence of QME in a broad coupling regime (\( 1/r \leq 0.15 \))~\cite{supp_cite}, demonstrating its robustness to weak integrability breaking. Consistently, level-spacing statistics confirm the emergence of quantum chaos with increasing long-range interaction strength (see Supplemental Fig. S9), corroborating that the system enters a regime with pronounced integrability breaking.

    \textit{Suppression of QME in intermediate coupling regime}---A primary experimental challenge lies in achieving precise control over the system Hamiltonian to investigate the sensitivity of QME to thermalization processes. We address this challenge in the intermediate coupling regime (\( r \approx 1 \)), where the interaction strength is tuned \(\bar{g}/2\pi \approx -2 \, \text{MHz}\)~\cite{supp_cite}. In this configuration, we prepare the same tilted initial states as in the \(r\approx10\) case. Here, thermalization governs the system dynamics, driving the subsystem towards thermal Gibbs ensembles—a process inherently linked to $ U(1) $ symmetry restoration in the Hamiltonian $ H $.
    
    Figure~\ref{fig2}(b) displays the EA dynamics for subsystem \(A\), revealing a key departure from the strong short-range regime: states with greater initial symmetry breaking exhibit slower symmetry restoration. Therefore, the dynamical crossover vanishes, resulting in the suppression of QME. Prior studies revealed a positive correlation between the speed of symmetry restoration and the thermalization rate~\cite{random_circuits_2024}. To investigate the relationship experimentally, we compute the EE derived from reduced density matrices, as shown in the inset of Fig.~\ref{fig2}(b). 
    The EE for both initial states increases linearly before saturating and approaches the Page value, signaling thermalization. Moreover, EA exhibits an initial near-linear decrease, subsequently approaching zero gradually, mirroring the dynamical evolution of the EE. 
    Meanwhile, the state with a smaller tilt angle exhibits faster entropy growth---characteristic of accelerated thermalization---which suppresses the emergence of QME. Therefore, our findings not only demonstrate that flexible modulation of coupling configuration enables the manipulation of QME, but also elucidate an intimate link between the restoration of system symmetry and thermalization dynamics.

	\textit{Reemergence of QME}---Building on the aforementioned insights, an intriguing question arises: can we leverage the flexibility of our system---such as varied on-site potentials and diverse initial states---to recover QME? Previous theoretical studies have shown that the suppression of thermalization and the breakdown of ergodicity lead to the emergence of QME for arbitrary tilted product states~\cite{QME_MBL}. 
    
    Inspired by these findings, we introduce the position-dependent on-site potential \(h_j/2\pi = W(7.5-j)\) for \(Q_j\) to break the ergodicity of our system. To quantify the suppression of ergodicity, we compute the average level-spacing ratio \(\langle r \rangle\)~\cite{supp_cite,2013_PRL_Level_Ratio}. The ergodic phase is characterized by Wigner-Dyson statistics (\(\langle r \rangle \approx\) 0.531), while the localized phase exhibits Poisson statistics (\(\langle r \rangle \approx\) 0.386)~\cite{2019_PNAS_mbl,2020_PRB_mbl}. In our system, as \(W\) increases, \(\langle r \rangle \) approaches 0.386 (see the inset of Fig.~\ref{fig2}(c)), signifying suppressed ergodicity and emergent localization-like features within experimentally accessible finite-size regimes. The EA dynamics for \(W = 6 \, \text{MHz}\) in Fig.~\ref{fig2}(c) reveal overall slower symmetry restoration than the on-resonance case. Meanwhile, the suppression of symmetry restoration exhibits tilt-angle dependence, with larger angles $\theta$ showing weaker suppression, thereby regenerating a dynamical crossover that recovers QME. Long-time EA simulations within the accessible \(W\) range corroborate this mechanism~\cite{supp_cite}: as \(W\) increases, the late-time EA of \( \ket{\theta = \pi/4}_N \) rises while \( \ket{\theta = \pi/2}_N \) remains nearly constant. This phenomenon results in $\Delta S_A(\pi/4,t\to \infty)>\Delta S_A(\pi/2,t\to \infty)$, guaranteeing crossover emergence in our experimental regimes.

\begin{figure}[htbp]
         \centering
     	\includegraphics[width=1.02\linewidth]{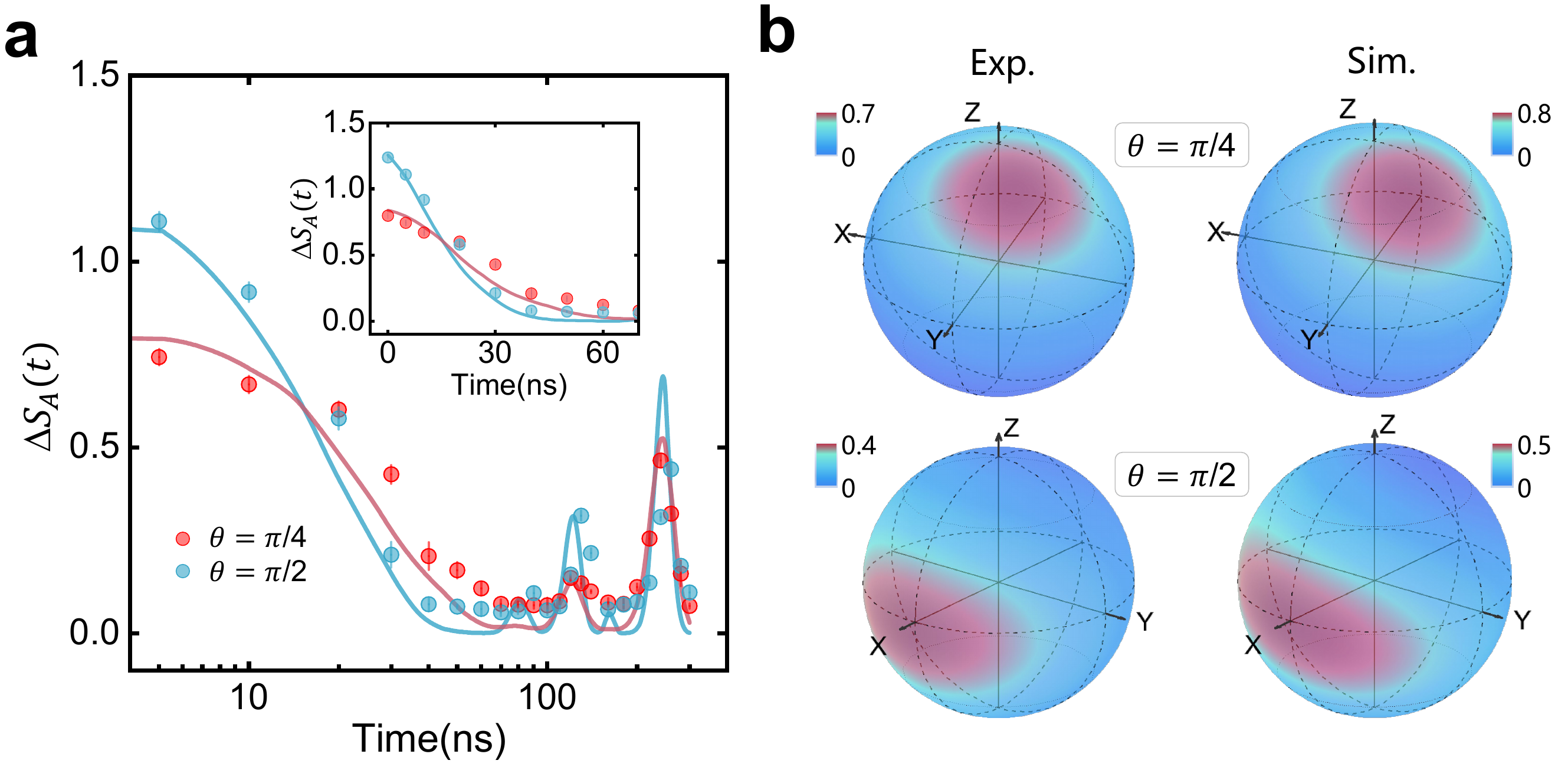}
     	\caption{EA dynamics in the intermediate coupling regime ($ r \approx 1 $) with tilted ferromagnetic states. (a) EA dynamics for two tilted ferromagnetic states with qubits on resonance. The inset shows EA dynamics with a linear time axis. Numerical simulations with decoherence are shown as solid lines, and experimental data are points. Error bars indicate the standard deviation of the experimental data. (b) Experimental and numerical data of quasidistribution function in spherical coordinates with evolution time \(t~\approx30\) ns.}
	\label{fig3}
     \end{figure}

    An alternative strategy to recover the QME is to quench from tilted ferromagnetic states. In this case, by introducing the collective spin operators \( \boldsymbol{S}\), the early-time dynamics can be mapped onto the LMG model, a paradigmatic system for studying dynamical phase transitions and many-body entanglement~\cite{2020_SA_xu}. Figure~\ref{fig3}(a) shows the EA dynamics quenched from tilted ferromagnetic states with $\theta = \pi/4$ and $\pi/2$, respectively. Notably, increased tilt angle $\theta$ enhances early-time symmetry restoration, thereby recovering QME with an obvious crossover. The subsystem dynamics are confined to the $S_A = N_A/2$ subspace, leading to the time-independence of the projected density matrix $\rho_{A,Q}$.
    This directly yields \(\Delta S_A(\theta,t) = C(\theta) - S_A(\theta,t)\), with $C(\theta)$ as an initial-state-dependent constant~\cite{supp_cite}. Therefore, the symmetry restoration rate precisely tracks the growth rate of EE. However, the QME cannot be detected by EE alone, because states with different tilt angles $\theta$ share the same initial value \(S_A(\theta,0) = 0 \). In contrast, EA shows crossover arising from the monotonic increase of both $C(\theta)$ and the initial entropy-growth rate with $\theta$. For the ideal LMG model, we derive exact solutions for the EA dynamics, further exploring the relationship among the crossover behavior, entropy type, and system size \cite{supp_cite}.

    Although the LMG model is exactly solvable, its symmetry restoration dynamics cannot be explained by quasiparticle propagation. Instead, the process is best visualized through the quasidistribution function $Q(\theta,\phi) = \langle \theta,\phi | \rho_A | \theta,\phi \rangle$, where $|\theta,\phi\rangle$ represents a coherent spin state~\cite{2019_science}.
    The dynamical flattening of the azimuthal ($\phi$-direction) distribution directly reflects the restoration of $U(1)$ rotational symmetry. As demonstrated in Fig.~\ref{fig3}(b), the distribution for $\theta = \pi/2$ exhibits faster flattening compared to $\theta = \pi/4$, providing a clear signature of the QME. In addition, recent studies employing time-dependent spin wave analysis show that $\Delta S_A$ directly corresponds to semi-classical spin fluctuations in the azimuthal direction~\cite{QME_long_range_spin}, which agrees well with the $Q$-function description of the dynamics. Furthermore, the collective spin dynamics display characteristic periodic behavior: the EA oscillates with period $T_{\text{EA}} = \pi/\bar{g}$~\cite{supp_cite}. 
     
\begin{figure}[htbp]
         \centering
     	\includegraphics[width=0.95\linewidth]{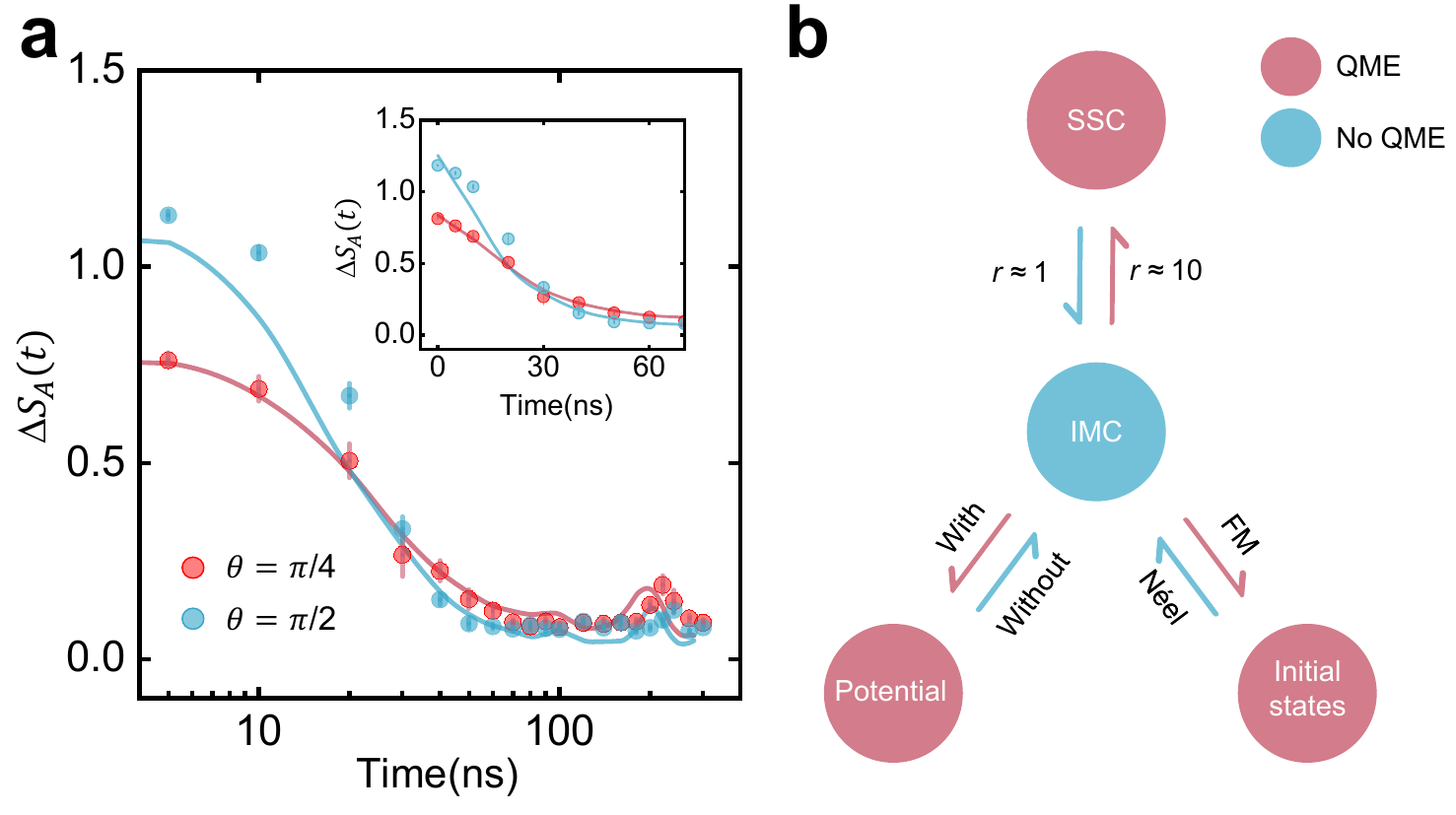}
     	\caption{(a) EA dynamics for two tilted ferromagnetic states in the presence of on-site disorder with $ r \approx 1 $. Data points denote experimental results with 6 disorder realizations, while solid lines represent theoretical results averaged over 100 disorder realizations. Error bars indicate the standard error of the experimental results. (b) Multidimensional modulation of QME. The QME emerges in strong short-range coupling (SSC) regime, is suppressed in intermediate coupling (IMC) regime, and reemerges under on-site potentials or from quenches of tilted ferromagnetic (FM) states.}

	\label{fig4}
     \end{figure}

    To suppress oscillations and unambiguously demonstrate the QME, we introduce on-site disorder to break spin equivalence. Figure~\ref{fig4}(a) shows that while periodic revivals are suppressed, the EA crossover persists under disorder uniformly distributed in \( [-14\bar{g}, 14\bar{g}]\). Despite slowing the relaxation and slightly prolonging the EA crossover time, disorder does not eliminate the QME.
    Our observation of QME with strong disorder demonstrates its robustness for tilted ferromagnetic initial states in the intermediate-coupling regime. Previous studies of power-law XX spin chains have reported that the crossover disappears within the experimental timescales under comparable disorder strengths~\cite{QME_trapped_ion}. These studies considered the Hamiltonian in Eq.~\eqref{hamiltonian} with \( g_{ij} = g_0 / |i - j|^\alpha \), where \( g_0 \) is a constant and \( \alpha \approx 1 \). In contrast, the Hamiltonian $H$ with \( r = 1 \) corresponds to exponent \( \alpha = 0 \) for a power-law XX spin chain, which exhibits stronger long-range interactions than the \( \alpha \approx 1 \) case. Therefore, our results reveal that stronger long-range interactions facilitate symmetry restoration, leading to more rapid EA crossovers within the experimental timescale.

	\textit{Conclusion and outlook}---In summary, we report the experimental realization of multidimensional modulation of the QME within an isolated quantum system, achieved using a quantum processor that features a fully connected architecture with flexible Hamiltonian parameters. Our study traces the QME emergence, suppression, and resurgence, providing a comprehensive framework for its control (see Fig.~\ref{fig4}(b)). Looking ahead, our work establishes the QME as a general and controllable feature of non-equilibrium quantum dynamics. The demonstrated modulation of QME suggests a promising route for accelerating equilibration, which is highly relevant to emerging protocols for efficient state preparation and qubit reset~\cite{lejeune2026acceleratingqubitresetmpemba,solanki2025universalrelaxationspeedupopen}. Future investigations could explore QME in prethermalization regimes~\cite{2025_periodicallydrivenEA,prether_IME_arxiv} or under alternative symmetries, such as \( \mathbb{Z}_2 \) and \( SU(2) \), revealing deeper insights into this phenomenon. Furthermore, probing EA dynamics under symmetry-breaking Hamiltonians presents an intriguing avenue to deepen our understanding of the interplay between symmetry and thermalization~\cite{2025_QME_no_sym,Breaking_Hamitonion_zhang}.
	
	~\\
	\textit{Acknowledgements}---We thank Zheng-Hang Sun for helpful discussions. This work was supported by National Natural Science Foundation of China (Grants Nos.~12447184, 12404578, 92265207, 92476202, 92365301, T2121001, 12204528, T2322030, 12247168), Innovation Program for Quantum Science and Technology (Grant No.~2021ZD0301800), Beijing Nova Program (Nos.~20220484121, 20240484652), Beijing National Laboratory for Condensed Matter Physics (2024BNLCMPKF022), Young Elite Scientists Sponsorship Program of the Beijing High Innovation Plan (Grant No.~20250945), and Beijing Natural Science Foundation (Grant No. 1262048). The device used in this work was made at the Nanofabrication Facilities at Institute of Physics, CAS in Beijing.

	~\\	
	\textit{Data availability}---The data that support the findings of this article are not publicly available. The data are available from the authors upon reasonable request.
	
	\bibliography{main.bib}

\onecolumngrid
\section*{End Matter}
\twocolumngrid
\textit{Measurement of Entanglement Asymmetry}---To extract EA, we reconstruct the density matrix $\rho$ using QST. 
The density matrix of an $n$-qubit subsystem is expanded in the Pauli operator basis as
\begin{equation}
\rho = \frac{1}{2^n} \sum_{\alpha} \mathrm{Tr}(\rho \sigma^{\alpha}) \, \sigma^{\alpha},
\tag{A1}
\end{equation}
where
$ \sigma^{\alpha} = \sigma_1^{\alpha_1} \otimes \sigma_2^{\alpha_2} \otimes \cdots \otimes \sigma_n^{\alpha_n} (\alpha_i \in \{I, X, Y, Z\}) $
denotes all possible tensor products of Pauli operators.
For an $n$-qubit subsystem, the Pauli expansion requires a $4^n$ operator basis, but only $3^n$ measurement bases are experimentally required, since the identity operator does not correspond to an independent measurement axis. In this work, for $n=3$, we implement $3^n = 27$ sets of measurement bases, each yielding $2^n = 8$ projective outcomes. By measuring expectation values along all relevant Pauli bases, the density matrix can be reconstructed.

Our superconducting processor natively supports measurement in the computational $Z$ basis. To access other Pauli bases, we apply single-qubit rotations $U$ prior to readout. For example, a $\pi/2$ rotation about the $X$ axis allows measurement in the $\sigma^Y$ basis. For pre-measurement rotation $U$, the state transforms as $\rho_{\mathrm{fin}} = U \rho_{\mathrm{ini}} U^\dagger$, with matrix elements
$\rho^{\mathrm{fin}}_{ij}
=
\sum_{k,l}
U_{ik}\,
\rho^{\mathrm{ini}}_{kl}\,
U^{*}_{jl}.$
The measurement outcomes correspond to the diagonal elements of 
$\rho^{\mathrm{fin}}$, forming a probability vector
$\vec{P} = \operatorname{diag}\!\left(\rho^{\mathrm{fin}}\right).$
Vectorizing the density matrix leads to the linear relation
$\tilde{U}\,\vec{\rho} = \vec{P},$
where
$\tilde{U}_{i,(k,l)} = U_{ik}\,U^{*}_{il}$
represents the superoperator form of $U$.

After collecting data from all $3^n$ Pauli measurement settings, we construct an overdetermined linear system and solve it using least-squares estimation. Since $\rho$ has $4^n - 1$ independent real degrees of freedom, this approach exploits the redundancy of the experimental dataset, which comprises $6^n$ total outcomes with $3^n(2^n-1)$ independent degrees of freedom. This overdetermined nature effectively suppresses readout noise and statistical fluctuations, providing a more robust estimate of $\rho$ for $n>1$. To ensure the reconstructed state is physically valid, we enforce Hermiticity, unit trace, and positive semi-definite constraints. The results were further validated using maximum likelihood estimation, showing high consistency.

To obtain the EA, we project the density matrix $\rho_A$ of subsystem $A$ into symmetry-resolved charge sectors, yielding the projected state $\rho_{A,Q}$, as illustrated in Fig.~\ref{fig1}(b). Through classical post-processing, we compute the von Neumann entropy $S(\rho_A)$ and the symmetry-projected entropy $S(\rho_{A,Q})$. The EA is then evaluated according to Eq.~\eqref{EA}.

%
	
\end{document}